# Quantum teleportation of an elemental silicon nanophotonic CNOT gate

Kai-Chi Chang[1,†,*], Xiang Cheng[1,†,*], Felix Ribuot-Hirsch[1], Murat Can Sarihan[1], Yujie Chen[1], Jaime Gonzalo Flor Flores[1], Mingbin Yu[2,3], Patrick Guo-Qiang Lo[2,4], Dim-Lee Kwong[2], and Chee Wei Wong[1,*]

[1] Fang Lu Mesoscopic Optics and Quantum Electronics Laboratory, University of California, Los

[2] Institute of Microelectronics, Singapore 117685, Singapore

[3] State Key Laboratory of Functional Materials for Informatics, Shanghai Institute of Microsystem and Information Technology, and Shanghai Industrial Technology Research Institute, Shanghai 200050, China

[4] Advanced Micro Foundry, Singapore 117685, Singapore

† These authors contributed equally to this work.

* Corresponding author emails: uclakcchang@ucla.edu, chengxiang@ucla.edu, and cheewei.wong@ucla.edu

**Large-scale quantum computers possess the capacity to effectively tackle practical problems that can be insurmountable for classical computers. The main challenge in building these quantum computers is to realize scalable modules for remote qubits and entanglement. By assembling small, specialized parts into a larger architecture, the modular approach mitigates complexity and uncertainty. Such a distributed architecture requires non-local quantum gate operations between remote qubits. An essential method for implementing such operations, known as quantum gate teleportation, requires only local operations, classical communication, and shared entanglement. Till today, the quantum gate teleportation using a photonic chip has remained elusive. Here we experimentally demonstrate the quantum teleportation of an on-chip controlled-NOT (CNOT) gate, assisted with the scalable silicon chip platform, high-fidelity local quantum logic gates, linear optical components, post-selected entanglement, and coincidence measurements from photonic qubits. First, we measure and characterize our teleported chip-scale CNOT gate with an average truth table fidelity of 93.1 ± 0.3%. Second, for different input polarization states, we obtain an average quantum state fidelity of 87.0 ± 2.2% with our teleported on-chip CNOT gate. Third, we use our non-local CNOT gate for remote entanglement creation of four Bell states, with an average quantum state fidelity of 86.2 ± 0.8%. Fourthly, we fully characterize**

**our teleported on-chip CNOT gate with a quantum process fidelity 83.1 ± 2.0%, and an average non-local CNOT gate fidelity of 86.5 ± 2.2%. Our teleported photonic on-chip quantum logic gate could be extended both to multiple qubits and chip-scale modules towards fault-tolerant and large-scale distributed quantum computation.**

Quantum advantage of non-classical machines require coherent manipulation of a large number of qubits [1-9]. As the system size grows, the average distance between qubits increases, making it harder to perform high-fidelity logical operation among connected arbitrary qubits. To tackle this problem, quantum state teleportation has been proposed and realized in different physical systems [10-16]. However, the teleportation of quantum logic gate operations themselves are more challenging than teleportation of states due to the communications cost overhead. Indeed, by using quantum state teleportation, a two-qubit quantum logic gate between two parties consumes a minimum of two shared Einstein, Podolsky and Rosen (EPR) entangled qubits, four classical bits, and several local operations [17-19]. Quantum gate teleportation, on the other hand, provides a unique solution that enables non-local quantum logical gate between remote qubits, where the shared EPR entanglement eliminates the requirement for a direct quantum interaction [20-22]. This key solution enables a modular approach to large-scale quantum computing [23-27], because the small and spatially separated processors with available qubits can be arbitrarily connected with quantum links to carry out more complicated computations. Such quantum network approach has provided practical scaling methods for several promising candidate systems for the implementation of quantum computation, for example, in photonic systems [20-22, 28], trapped-ion platforms [29-32], and in superconducting hardware [33]. All of these prior works demonstrated the CNOT gate teleportation, owing to its capability to entangling different qubits at the proper settings [34-44], and such two-qubit gate has been proven to play a crucial role in entanglement distillation and purification [45-48]. The CNOT gate, when combining with single-qubit gates, enabling the generation of highly entangled state for one-way quantum computing [49-57]. To the best of our knowledge to date, although the photonic quantum state teleportation has been demonstrated on-chip [58] and from one chip to another chip [59], the quantum teleportation of the gate operation using a photonic chip, however, is lacking.

Here we experimentally realize a teleported on-chip CNOT gate between two remote nodes using a shared EPR pair of photonic qubits. This demonstration combines key elements for non-local quantum logic gate with flying qubits, including our complementary metal-oxide-semiconductor (CMOS) compatible silicon chip [60], linear optical manipulations and high-fidelity local quantum logic gates, the assistance of post-selected entanglement generation and coincidence counting measurements. Here we teleport a local on-chip CNOT gate, which acts on polarization and path qubits of two local qubits, to a non-local CNOT gate, acting on polarization qubits of two remote qubits. First, we employ non-local truth table measurements to describe our teleported chip-scale CNOT gate, with an average gate fidelity of 93.1 ± 0.3%. Second, we obtain an average fidelity of 87.0 ± 2.2% for various polarization states using our teleported chip-scale CNOT gate. In the third step, our non-local CNOT gate is utilized to remotely create entanglement for all four Bell states, yielding an average quantum state fidelity of 86.2 ± 0.8%. Finally, we comprehensively characterize the logical quantum process of our on-chip CNOT gate through teleported measurements, revealing a complete quantum process fidelity of 83.1 ± 2.0%. Therefore, we demonstrate the first quantum teleportation of a chip-scale CNOT gate, exhibiting a mean gate fidelity of 86.5 ± 2.2%, averaged across various input states. Our on-chip CNOT gate device has a size of 50×50 μm$^2$, which is promising for replacing bulky free-space optics components with CMOS integration technology, this could further increase the number of connected modules and qubits for distributed quantum computation. Our teleported on-chip CNOT gate operation can be incorporated in a larger modularized quantum computer that is distributed through optical fibers over longer distances. The demonstration of teleportation of an on-chip CNOT gate here, apart from its fundamental interest, is a significant step towards the realization of quantum network, and distributed quantum computing using photonic qubits.

## Results

**Silicon chip CNOT gate teleportation realization**

A distributed quantum computer relies on a network of local nodes that establish communication with each other in both quantum and classical channels (Fig. 1A). Each node represents a compact quantum processor comprising two finely-optimized subsystems. Now that we have two parties, Alice and Bob, each having two qubits 1, 2 and 3, 4, each with quantum information. We have a shared EPR state $|\Phi\rangle_{23} = (|00\rangle_{23} + |11\rangle_{23})/\sqrt{2}$ between the qubits 2 and 3, facilitating interactions between modules by acting as communicators. Each module functions independently as a local node, capable of executing intra-module operations proficiently. Inter-module operations between the two entities are made possible by distributing entanglement among the communication qubits. As a consequence of the isolation between modules, traditional approaches relying on direct quantum interactions cannot be employed to implement multi-qubit operations in the network architecture. Instead, quantum state teleportation is developed for this purpose [12-19]. This protocol relies on shared entanglement as a valuable resource, along with deterministic local operations and classical communication between two systems. The combination of these elements forms the distinctive characteristics of teleportation-based protocols, where information is transmitted through quantum and classical channels. Building upon this technique, quantum gate teleportation achieves a unitary gate operation between two unknown states, eliminating the requirement for direct interaction between non-local qubits [20-22]. First, we briefly explain the fundamental idea of the teleportation of our chip-scale quantum gate operation in photonic platform. The input state of the qubit 1, 4 can be arbitrary expressed as $|\Psi\rangle_{14} = A_1|00\rangle + A_2|01\rangle + A_3|10\rangle + A_4|11\rangle$, where $A_1$ to $A_4$ is the arbitrary normalization value, then, we perform deterministic local gate operations between qubit 1, 2, and qubits 3, 4, respectively. Through shared EPR entanglement between qubit 2 and 3 and classical communications, we can teleport a local gate between qubits 1 and 2 to remote qubits 1 and 4 through the process: $C_{34}C_{12}(|\Psi\rangle_{14} \otimes |\Phi\rangle_{23}) = |0\,+\rangle_{23} \otimes C_{14}|\Psi\rangle_{14} + |0\,-\rangle_{23} \otimes (\sigma_1^Z)C_{14}|\Psi\rangle_{14} + |1\,+\rangle_{23} \otimes (\sigma_4^X)C_{14}|\Psi\rangle_{14} + |1\,-\rangle_{23} \otimes (-\sigma_1^Z\sigma_4^X)C_{14}|\Psi\rangle_{14}$, where $\sigma_1^Z$, $\sigma_4^X$ are the Pauli operators acting on the corresponding qubits, $|\pm\rangle_3 = (|0\rangle_3 \pm |1\rangle_3)/\sqrt{2}$, with $C_{34}$ and $C_{12}$ the local gate operations. The entire process therefore teleports our quantum on-chip CNOT gate from local qubits 1, 2 to remote qubits 1, 4 ($C_{12} \rightarrow C_{14}$). This technique has the minimum classical communication cost for a non-local quantum chip-scale CNOT gate compared to quantum state teleportation protocols [16, 30-33].

In this work we realize the quantum teleportation of our photonic silicon chip-scale CNOT gate. The design optimized map of our on-chip CNOT gate is given in our prior studies [60]. To achieve efficient chip-scale polarization splitting, a waveguide-to-waveguide gap width of 400 nm and a coupling length of 11.5 μm are implemented. The silicon waveguides are optimized with a thickness of 220 nm for C-band operation. Fig. 1B is the optical micrograph of the on-chip CNOT gate (size of 50×50 μm², excluding side couplers) by the integrated silicon nanophotonics polarized coupler. For the nanofabrication detail of our on-chip CNOT gate, see Methods in Supplementary Materials. The classical transmission for this chip is provided in our previous work [60], here we choose to use the best devices from same chipset that has high-fidelity two-qubit gate operations.

To successfully implement the experimental demonstration of teleporting our chip-scale CNOT gate, two key requirements must be met: preparing the resource for standard teleportation and enabling the execution of CNOT gates on local qubits. In our physical implementation, we leverage the advantages of the photonic platform, utilizing highly coherent and controllable components. Our measurement setup is presented in Fig. 1C. With the type-II spontaneous parametric down-conversion (SPDC) process, photon-pairs are generated in a post-selected polarization-entangled EPR state $(|HV\rangle + |VH\rangle)/\sqrt{2}$, where $|H\rangle$ and $|V\rangle$ stand for two orthogonal linear polarizations. Then, we use two polarization beam-splitters (fiber $PBS_2$ and $PBS_3$) to produce individually controllable path qubits for Alice and Bob. We assign the two paths after $PBS_2$ as $|0\rangle$ and $|1\rangle$, respectively. We use a fiber polarization controller ($FPC_3$) and a half-wave plate ($HWP_4$) to change the polarization from $|V\rangle$ to $|H\rangle$. By doing so, we can transfer the polarization-entanglement to path-entanglement with whole quantum state expressed as $|\Psi\rangle_{1234} = |H\rangle_1[(|00\rangle_{23} + |11\rangle_{23})/\sqrt{2}]|H\rangle_4$, before gate operations. Alice and Bob then carry qubits 1, 2 and 3, 4, respectively. The local on-chip CNOT gate $C_{12}$ is achieved by our on-chip silicon device, while the other local CNOT gate $C_{34}$ is realized in a free-space balanced Mach-Zehnder interferometer. For other experimental details, see Methods in Supplementary Materials. Subsequently, we prepare arbitrary polarization input states between remote qubits 1 and 4 to implement first quantum teleportation of an on-chip CNOT gate operation.

**Teleported chip-scale CNOT gate truth able and quantum state tomography**

Before performing non-local CNOT gate protocol, first we measure and optimize spectral, temporal, and polarization characteristics of our photon-pair source. By carefully sweeping the pump wavelength, we measure that our SPDC source has an ≈ 0.2 nm center-wavelength difference, operating at near-degenerate condition. We then proceed to measure the Hong-Ou-Mandel (HOM) two-photon interference to characterize our photon-pair source. We obtain a HOM visibility of 97.4 (92.2) ± 0.8% after (before) background subtraction. We also use HOM interference dip to confirm that the path-lengths in our free-space Mach-Zehnder interferometer is balanced. After these measurements, we generate polarization-entanglement via post-selection, with 98.5 (93.9) ± 1.0 % fringe visibility after (before) background subtraction. For polarization fringes after accidental subtraction, we obtain $S = 2.686 \pm 0.033$ to violate the classical limit by more than 20 standard deviations [35]. All these results are presented in Figures S.1 to S.3 of Supplementary Materials Section I. We next measure the truth tables for two local CNOT gates as illustrated in Fig. 2A and 2B. For our chip-scale CNOT gate, we directly prepare and send SPDC photons into dual input-output coupling systems, and insert a pair of half-wave plates (HWP$_2$ and HWP$_3$) on both input ports before the chip to prepare the $|H\rangle$ ($|0\rangle$) and $|V\rangle$ ($|1\rangle$) polarizations. We use two linear polarizers (P$_1$ and P$_2$) before output collimators to perform projection polarization measurements. In Fig. 2A, the local on-chip CNOT gate has an average gate fidelity of $97.9 \pm 0.3\%$, consistent with our prior studies [60]. As for the free-space CNOT gate truth table measurements, we use the SPDC source, removing the FPC$_5$, FPC$_6$, two fiber tunable delay lines, fiber BS$_2$, and connect the outputs of free-space CNOT gate directly to the two fiber benches. Input polarization states $|H\rangle$ ($|0\rangle$) and $|V\rangle$ ($|1\rangle$) can be prepared with HWP$_7$ and HWP$_8$ (while the HWP$_5$ is fixed at 45º to act as second local CNOT gate), and output polarization projection measurements is done with mounted polarization analyzers (P$_3$ and P$_4$). For Fig. 2B, the second local CNOT gate has an average gate fidelity of $98.1 \pm 0.3\%$. In both Fig. 2A and 2B, errors are given as black ranges and indicate the standard deviations in our measurement, and the expected output states for an ideal CNOT gate are shown as transparent bars. All the experimental data here are obtained without subtracting accidental coincidences.

With the successful demonstration of all the essential elements required for the implementation of the teleported on-chip CNOT gate, we proceeded to thoroughly characterize the non-local two-qubit chip-scale gate through a series of four analyses. In the following, we define the physical qubits $|H\rangle$, $|V\rangle$, $|D\rangle$, $|R\rangle$ in the logical basis as $|0\rangle$, $|1\rangle$, $|+\rangle = (|0\rangle + |1\rangle)/\sqrt{2}$, and $|i\rangle = (|0\rangle + i|1\rangle)/\sqrt{2}$ respectively. In the first analysis, we confirmed the functionality of the non-local gate by measuring quantum truth tables for the complete set of computational basis states. We prepared the input qubits for all four states $|00\rangle$, $|01\rangle$, $|10\rangle$, and $|11\rangle$ and operated the teleported chip-scale CNOT on them. Fig. 2C is the experimental teleported truth table of our on-chip CNOT gate. To assess the fidelity of the measured teleported CNOT truth table compared to the ideal one, we performed a calculation that quantifies the degree of similarity between the two: $F = (1/4)Tr\left(\frac{M_{exp}M_{ideal}^T}{M_{ideal}M_{ideal}^T}\right)$, where $M_{exp}$ and $M_{ideal}$ are the truth tables for experiment and ideal cases respectively. Here we optimized and obtained an average truth table fidelity $F$ of $93.1 \pm 0.3\%$. We note that the deviation from unity teleported fidelity arises from our chip-scale CNOT's finite polarization-extinction ratio, coupling difference between the $|H\rangle$ and $|V\rangle$ states [60], imperfections from free-space local CNOT gate, slight mismatch between signal and idler wavelength that causes non-ideal HOM interference visibility and polarization-entanglement quality (detailed in Figure S.1 to S.3 in Supplementary Materials Section I). Errors are given as black ranges and indicate the standard deviations in our measurement. The expected output states for an ideal CNOT gate are shown as transparent bars. All the data are obtained without subtracting accidental coincidences. The results obtained in this analysis offer initial qualitative verification of the teleported chip-scale CNOT gate's performance.

Subsequently, in the second step, we employ quantum state tomography to analyze the on-chip CNOT gate teleportation between remote qubits 1 and 4 (see Methods and Supplementary Materials Section II for more details), given different input states composed of $|0\rangle$, $|1\rangle$, $|+\rangle$, and $|i\rangle$. The quantum state fidelity can be expressed as: $F_s = (Tr(\sqrt{\sqrt{\rho_{mea}}\rho_{ideal}\sqrt{\rho_{mea}}}))^2$, where $\rho_{mea}$ and $\rho_{ideal}$ are the measured and ideal density matrices respectively. We represent all of the different input polarization states' reconstructed density matrices in Figures S.4 to S.7 in Supplementary Materials Section II. Overall, we measure and achieve an average quantum state fidelity of 87.0 ± 2.2% for 14 different input polarization states, namely, $|00\rangle$, $|01\rangle$, $|0+\rangle$, $|0i\rangle$, $|10\rangle$, $|11\rangle$, $|1+\rangle$, $|1i\rangle$, $|++\rangle$, $|+i\rangle$, $|i0\rangle$, $|i1\rangle$, $|i+\rangle$, and $|ii\rangle$ (the other 2 input polarization states $|+0\rangle$, $|+1\rangle$ are presented in the third analysis) with our teleported chip-scale CNOT gate. In addition to the errors stemming from imperfect experimental non-local truth table, these quantities encompass imperfections related to logical state preparation and decoding operations. The ideal values are indicated as transparent bars in the plots. The duration of coincidence counting for each experimental data for quantum state tomography results is 10 second, and all the data are obtained without subtracting accidental coincidences. Our quantum state tomography results in Supplementary Materials Section II are consistent with our non-local truth table measurements in Fig. 2C, further confirming the non-local gate operation between remote qubit 1 and 4 using our teleported chip-scale CNOT gate.

**Generation of logical bell states and quantum process tomography of an on-chip CNOT gate teleportation**

After verifying the non-local CNOT operation with our truth table and quantum state tomography measurements, we proceed to characterize its unique entangling properties. In our third analysis, we represent the unique quantum nature of our teleported on-chip gate by successfully creating entanglement between two distant logical qubits. To achieve this, we initialize the input remote qubits in the separable initial state $|+0\rangle$, $|+1\rangle$, $|-0\rangle$, and $|-1\rangle$, and perform the non-local gate operation respectively. Their ideal output state corresponds to four Bell states $|\Phi^+\rangle$, $|\Psi^+\rangle$, $|\Phi^-\rangle$, and $|\Psi^-\rangle$, respectively. We express the four maximally-entangled Bell states as $|\Phi^\pm\rangle = (|0\rangle|+\rangle \pm |1\rangle|-\rangle)/\sqrt{2}$, and $|\Psi^\pm\rangle = (|0\rangle|-\rangle \pm |1\rangle|+\rangle)/\sqrt{2}$, where $|-\rangle = (|0\rangle - |1\rangle)/\sqrt{2}$. We note that the input polarization states of $|+0\rangle$, $|+1\rangle$, combining with other 14 polarization states in Supplementary Materials Section II, forms a complete quantum state tomography measurements using $|0\rangle$, $|1\rangle$, $|+\rangle$, and $|i\rangle$ input states. Fig. 3A to 3D exhibit the real and imaginary parts of the reconstructed density matrices for all four generated Bell states $|\Phi^+\rangle$, $|\Psi^+\rangle$, $|\Phi^-\rangle$, and $|\Psi^-\rangle$ via our chip-scale CNOT gate teleportation. For these produced Bell states using a teleported on-chip CNOT gate, we achieve an average quantum state fidelity $F_s$ of $86.2 \pm 0.8\%$. The transparent bars indicate the ideal density matrices for the maximally-entangled Bell states. The duration of coincidence counting for each experimental data in Fig. 3 is 10 second, and all the data are obtained without subtracting accidental coincidences. Our results here confirm the non-local entangling functionality of our teleported chip-scale CNOT gate.

In the fourth analysis, we fully characterize the quantum logical process of the teleported on-chip CNOT gate. To implement this, we conduct a quantum process tomography on the two remote qubits. For an ideal quantum process tomography of a CNOT gate, we depict the real and imaginary parts of the quantum process using the Pauli basis. In this representation, $X$, $Y$, and $Z$ correspond to the Pauli matrices $\sigma_X$, $\sigma_Y$, and $\sigma_Z$, respectively. To ensure a quality quantum logical process tomography for our teleported chip-scale CNOT gate, our approach focuses on optimizing each density matrix of the quantum state tomography by carefully selecting appropriate input and output states. In this study, we employ a total of 256 measurement settings, with each setting having a counting duration of 10 seconds, to perform a comprehensive quantum process tomography. The fidelity of the quantum process can then be expressed as: $F_p =$

$$\left(Tr\left(\sqrt{\sqrt{\chi_{exp}}\chi_{ideal}\sqrt{\chi_{exp}}}\right)\right)^2,$$ where $\chi_{exp}$ and $\chi_{ideal}$ are the experimental and ideal process matrices respectively. From the experimental reconstruction, we extract a process fidelity $F_p$ of 83.1 ± 2.0%. By using a relation between the quantum process fidelity $F_p$ and the average gate fidelity $F_{avg}$, with $F_{avg} = \frac{dF_p+1}{d+1}$, and $d = 4$ [34], we obtain an average teleported on-chip CNOT gate fidelity $F_{avg}$ of 86.5 ± 2.2% without subtracting accidental coincidences. The successful implementation of on-chip CNOT gate teleportation is confirmed by the agreement between the experimentally reconstructed process (Fig. 4B) and the ideal process (Fig. 4A). Our work here provides an important step towards achieving distributed quantum computation in a chip-scalable platform using photonic qubits.

## Discussion and outlook

In summary, we have demonstrated a quantum gate teleportation with our integrated silicon chip device, linear optical operations and local quantum logic gates, shared EPR entanglement and post-selected coincidence measurements in a photonic platform. The quality of our teleported chip-scale CNOT gate is first characterized through non-local truth table measurements with an average gate fidelity of 93.1 ± 0.3%. In the second analysis, we achieve an average quantum state fidelity of 87.0 ± 2.2% for different polarization states after our non-local CNOT gate operation. Third, we perform quantum state tomography to analyze the generation of logical Bell states after an on-chip CNOT gate teleportation, with an average quantum state fidelity of 86.2 ± 0.8%. Finally, we fully characterize the quantum logical process of our teleported on-chip CNOT gate with a quantum process fidelity 83.1 ± 2.0%. We therefore demonstrate the first teleportation of the chip-scale CNOT gate operation with a mean gate fidelity of 86.5 ± 2.2%, averaged over all the different input states. The integration of several operations in quantum gate teleportation, including CMOS-compatible silicon chip, high-fidelity quantum logic gates, efficient generation and manipulation of photonic qubits in optical link, will be essential for building large-scale quantum computers based on entangled photons. The protocol employed in our work for the teleported on-chip CNOT gate is one example from a broad range of two-qubit operations that can be achieved utilizing similar resources. These non-local quantum logic gates play a crucial role as fundamental building blocks for modular architectures. Our future endeavors will involve demonstrating non-local teleported gates using multiple chip-scale qubit modules that operate independently, necessitating remote entanglement and the assistance of photonic qubits in the telecommunication band. Additionally, our protocol can leverage other approaches, such as deterministic [8, 22, 30, 31, 33] and fault-tolerant [9, 32] schemes, and can benefit from entanglement purification protocols [47, 48]. When combining with non-destructive measurement schemes [61-65], our current implementation has the potential for future adaptation to incorporate more intricate schemes that generate deterministic remote entanglement. Such advancements will be crucial for achieving a scalable and modular qubit architecture. By combining our findings with recent advancements in photonic quantum computing [3, 4, 7, 10, 11], it becomes possible to synergize these technologies and facilitate the development of fault-tolerant and scalable modular quantum computing via photonic qubits.

## DATA AVAILABILITY

The datasets generated and analyzed during this study are available from the corresponding authors upon reasonable request. Source data are provided with this paper.

## ACKNOWLEDGEMENTS

The authors acknowledge discussions with Lan-Tian Feng, Xifeng Ren, Serdar Kocaman, Kangdi Kerry Yu, Sophi Chen Song, Hsiao-Hsuan Chin, and Zhenda Xie, and discussions on the superconducting nanowire single-photon detectors with Vikas Anant. This study is supported by the Army Research Office Multidisciplinary University Research Initiative (W911NF-21-2-0214) and the National Science Foundation (QII-TAQS 1936375 and QuIC-TAQS 2137984).

## AUTHOR CONTRIBUTIONS

K.-C.C., and X.C. developed the idea and designed the experiment. K.-C.C., X.C., F.R-.H., and Y.C. conducted the measurements. K.-C.C., F.R-.H., and J.G.F.F. contributed to the data analysis. M.C.S. performed the design and simulation of the device. M.Y., P.G-.Q.L., and D.-L.K. performed the device nanofabrication. X.C., F.R-.H., M.C.S., and C.W.W. supported and discussed the studies. K.-C.C., X.C., and C.W.W. prepared the manuscript. All authors contributed to the discussion of the manuscript.

## SUPPLEMENTARY MATERIALS

Science Advances/sciencemag.org/content/tbd

    Methods

Supplementary Text

Figure S.1 to S.7

References (*66-81*)

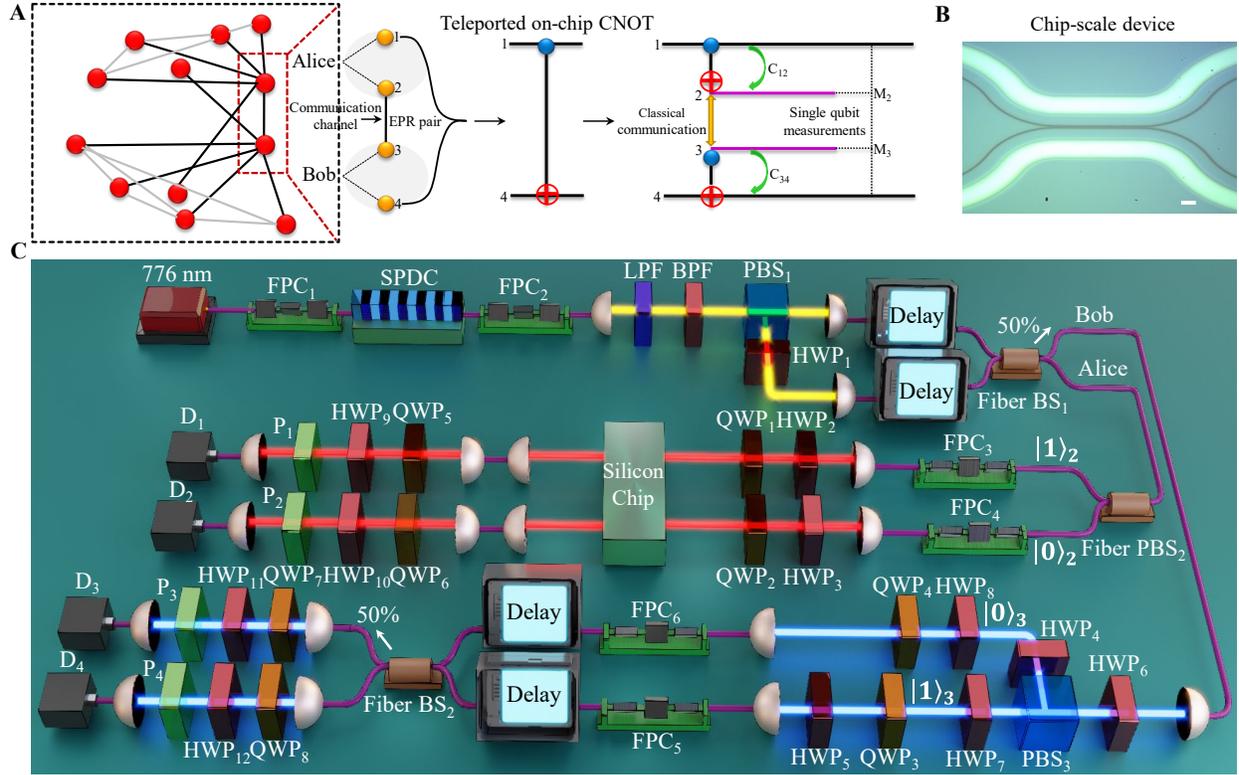

**Fig. 1. Quantum teleportation of an on-chip CNOT gate realization scheme. A,** Network overview of the distributed quantum computation. Alice and Bob each has two-qubit (1 to 4, in red circles), sharing an EPR pair between them. Coupling between Alice and Bob is generated through a communication channel. Their two-qubit comprises of a photonic quantum processor that is capable of high-fidelity operations among data qubits (1 and 4) and communication qubits (2 and 3). We then show the teleported CNOT logic between Alice and Bob, which requires (1) EPR entanglement between qubit 2 and 3 (orange arrow), (2) local CNOT operations (green arrows: $C_{12}$ and $C_{34}$), (3) Single-qubit measurement of qubit 2 and 3 in proper basis ($M_2$ and $M_3$), and (4) classical communication between Alice and Bob. **B,** Optical micrograph of the on-chip CNOT gate by the integrated silicon photonics polarized coupler. Scale bar: 2 $\mu$m. **C,** Experimental setup for quantum teleportation of an on-chip CNOT gate, which includes three main parts, 1. The post-selected polarization entanglement generation; 2. The chip-scale CNOT gate prepared for teleportation in Alice; 3. The free-space two-qubit setup prepared by Bob for teleporting Alice's on-chip CNOT logic operation. $|0\rangle_2$, $|1\rangle_2$, $|0\rangle_3$, and $|1\rangle_3$ are for the notation of path qubits after post-selected polarization entanglement generation. Our integrated CNOT gate is a silicon waveguide device realized by a polarized directional coupler. FPC, fiber polarization controller; LPF, long-pass filter; BPF, band-pass filter; PBS, polarization beam-splitter; BS, beam-splitter; HWP, half-wave plate; QWP, quarter-wave plate; P: linear polarizer; D, detector; TCSPC, time-correlated single-photon counting.

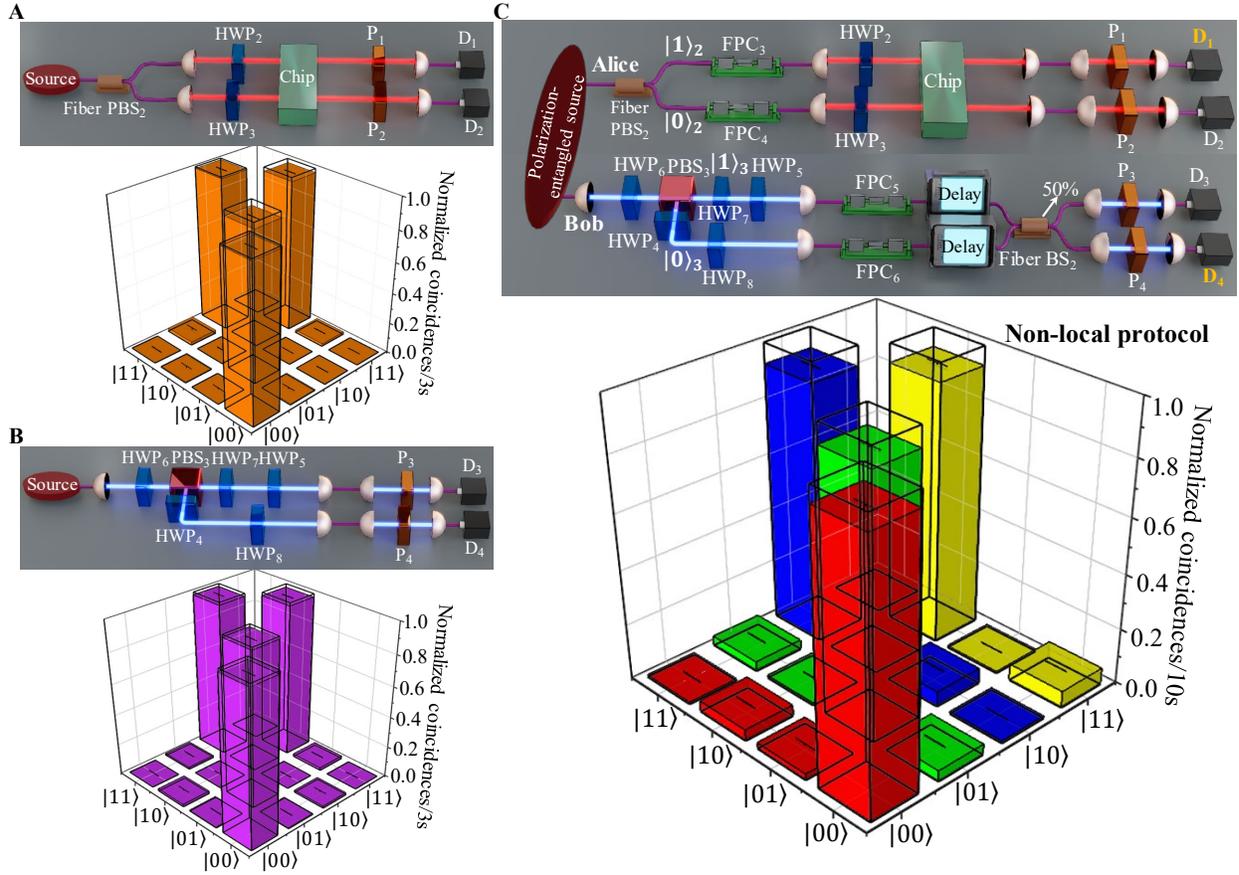

**Fig. 2. Teleported truth table of an on-chip CNOT gate and local CNOT truth tables. A,** Simplified experimental setup (detail in Fig. 1C) for truth table measurements of a chip-scale CNOT gate. The local CNOT gate has an average gate fidelity of 97.9 ± 0.3%. **B,** Free-space CNOT gate truth table and its measurement setup. The second local CNOT gate has an average gate fidelity of 98.1 ± 0.3%. Errors are given as black ranges and indicate the standard deviations in our measurements. In both Fig. 2A and 2B, errors are given as black ranges and indicate the standard deviations in our measurement, and the expected output states for an ideal CNOT gate are shown as transparent bars. **C,** Schematic experimental setup for non-local truth table measurements between detector 1 and 4 (highlighted in orange color). Below is the non-local protocol for teleporting a chip-scale CNOT gate. The non-local CNOT gate has an average truth table fidelity of 93.1 ± 0.3%. Errors are given as black ranges and indicate the standard deviations in our measurement. The expected output states for an ideal CNOT gate are shown as transparent bars. All the experimental data here are obtained without subtracting accidental coincidences.

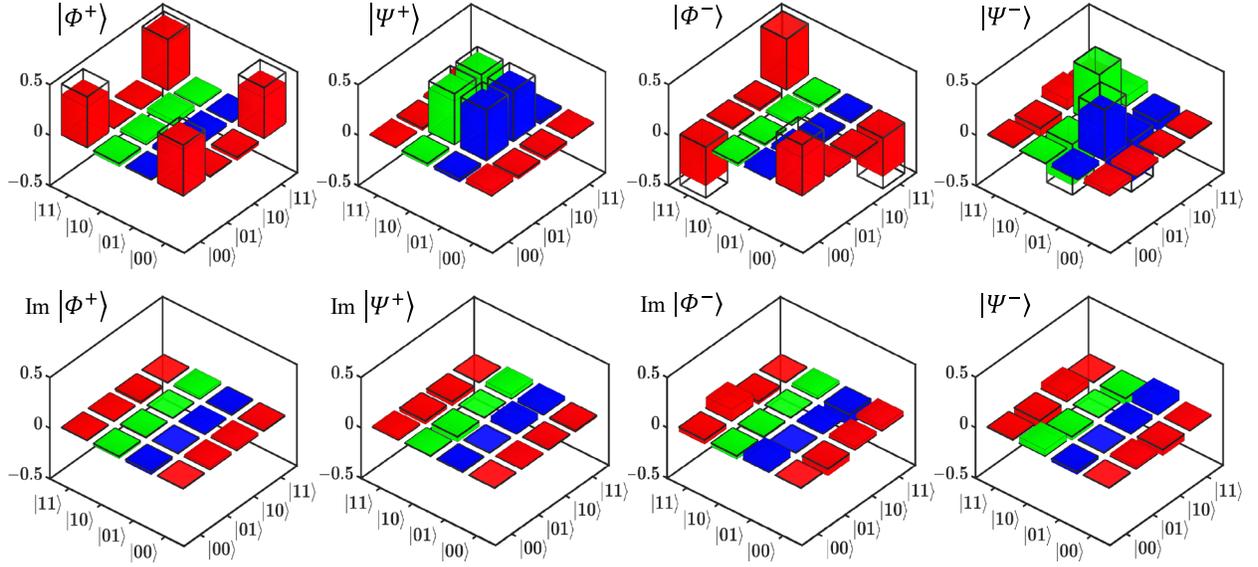

**Fig. 3. Reconstructed density matrices for the generated Bell states after a chip-scale CNOT gate teleportation.** Produced Bell states after an on-chip CNOT gate teleportation. We show both the real and imaginary parts of the reconstructed density matrices for four Bell states with four different input polarization states of **A,** $|+0\rangle$, **B,** $|+1\rangle$, **C,** $|-0\rangle$, and **D,** $|-1\rangle$, respectively. The outputs states are measured by 16 different output combinations of $|0\rangle$, $|1\rangle$, $|+\rangle$, and $|i\rangle$. For these produced Bell states using a teleported on-chip CNOT gate, we achieve an average quantum state fidelity of $86.2 \pm 0.8\%$. The transparent bars indicate the ideal density matrices for the maximally-entangled Bell states. The duration of coincidence counting for each experimental data in Fig. 3 is 10 second, and all the data are obtained without subtracting accidental coincidences.

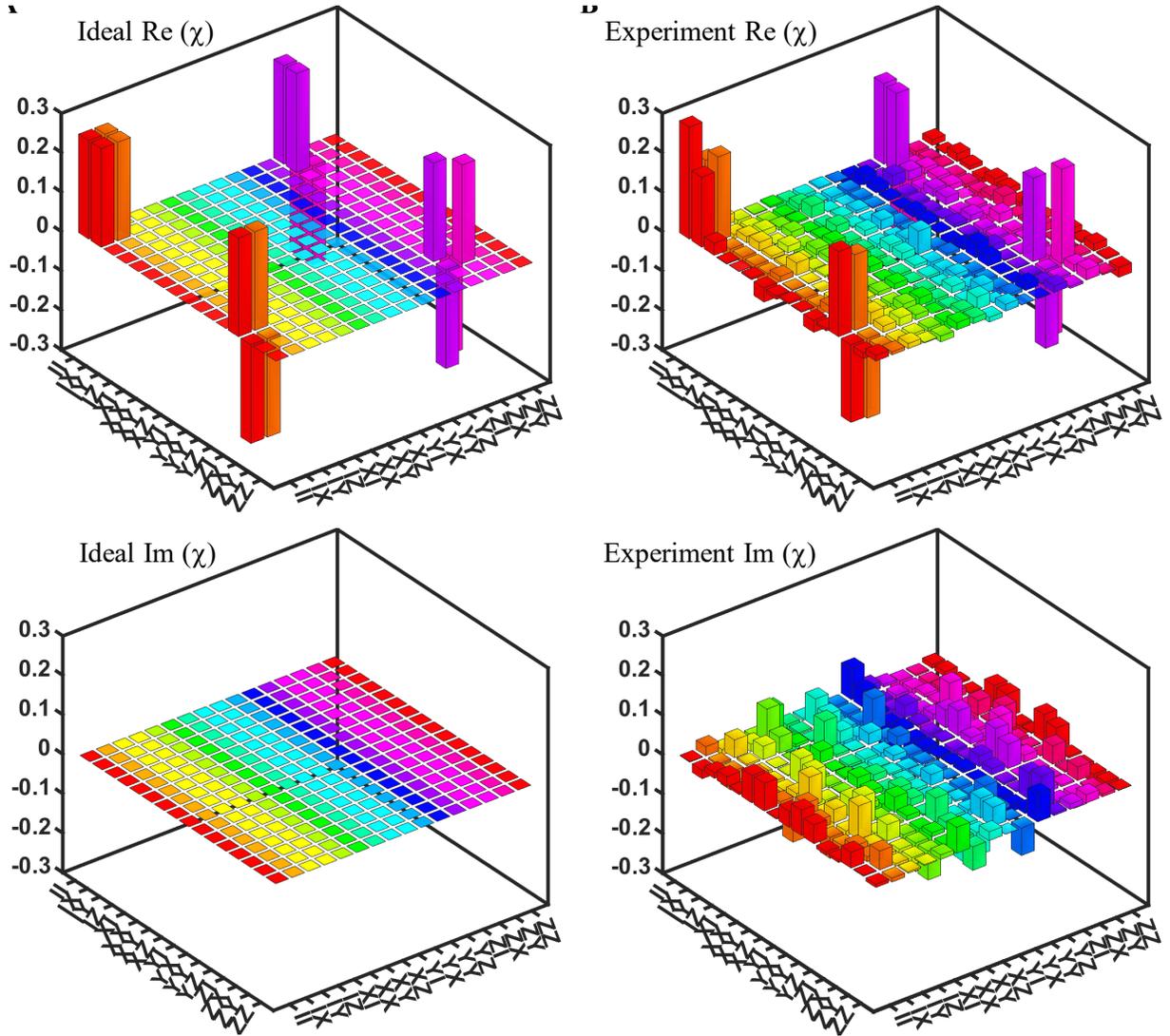

**Fig. 4. Teleported quantum process tomography of an on-chip CNOT gate. A,** Ideal quantum process tomography of the CNOT gate. We represent both the real and imaginary part of quantum process tomography of CNOT gate in the Pauli transfer representation, here, *X*, *Y*, and *Z* is the Pauli matrices $\sigma_X$, $\sigma_Y$, and $\sigma_Z$, respectively. **B,** Experimental reconstructed quantum process tomography of a teleported chip-scale CNOT gate. The teleported on-chip CNOT gate process fidelity is measured to be $83.1 \pm 2.0\%$, which corresponds to the average teleported on-chip CNOT gate fidelity of $86.5 \pm 2.2\%$. Agreement between the experimentally reconstructed (Fig. 4B) and ideal (Fig. 4A) processes indicates the successful implementation of an on-chip CNOT gate teleportation. The duration of coincidence counting for each experimental data in Fig. 4B is 10 second, and all the data are obtained without subtracting accidental coincidences.